\documentclass{article}

\usepackage[left=3cm, right=3cm, top=2cm, bottom = 2cm]{geometry}

\usepackage{lmodern}  
\usepackage[utf8]{inputenc}
\usepackage[T1]{fontenc}
\usepackage[UKenglish]{isodate}

\usepackage{amsmath}
\usepackage{graphicx}
\usepackage{xcolor}
\usepackage{tikz}
\usepackage{textcomp}

\usepackage{cite}

\usepackage{scalerel}
\usepackage{tikz}
\usetikzlibrary{svg.path}

\definecolor{orcidlogocol}{HTML}{A6CE39}
\tikzset{
  orcidlogo/.pic={
    \fill[orcidlogocol] svg{M256,128c0,70.7-57.3,128-128,128C57.3,256,0,198.7,0,128C0,57.3,57.3,0,128,0C198.7,0,256,57.3,256,128z};
    \fill[white] svg{M86.3,186.2H70.9V79.1h15.4v48.4V186.2z}
                 svg{M108.9,79.1h41.6c39.6,0,57,28.3,57,53.6c0,27.5-21.5,53.6-56.8,53.6h-41.8V79.1z M124.3,172.4h24.5c34.9,0,42.9-26.5,42.9-39.7c0-21.5-13.7-39.7-43.7-39.7h-23.7V172.4z}
                 svg{M88.7,56.8c0,5.5-4.5,10.1-10.1,10.1c-5.6,0-10.1-4.6-10.1-10.1c0-5.6,4.5-10.1,10.1-10.1C84.2,46.7,88.7,51.3,88.7,56.8z};
  }
}

\newcommand\orcidicon[1]{\href{https://orcid.org/#1}{\mbox{\scalerel*{
\begin{tikzpicture}[yscale=-1,transform shape]
\pic{orcidlogo};
\end{tikzpicture}
}{|}}}}

\usepackage[xindy]{glossaries}

\newacronym{ltl}{LTL}{Linear-time Temporal Logic}

\newacronym{ctl}{CTL}{Computation Tree Logic}

\newacronym{ptl}{PTL}{Probabilistic Temporal Logic}

\newacronym{pctl}{PCTL}{Probabilistic Computation Tree Logic}

\newacronym{pctl*}{PCTL*}{Probabilistic Computation Tree Logic*}

\newacronym{tctl}{TCTL}{Timed Computation Tree Logic}

\newacronym{stl}{STL}{Signal Temporal Logic}

\newacronym{ltl-x}{LTL-X}{a version of Linear Time Logic without the `next' operator}

\newacronym{mtl}{MTL}{Metric Temporal Logic}

\newacronym{iactlk-x}{IACTLK-X}{a fragment of the temporal-epistemic logic CTLK, without the `next' operator}

\newacronym{mucalc}{$\mu$-calculus}{a strict superset of other temporal logics}

\newacronym{fotl}{FOTL}{First-Order Temporal Logic}

\newacronym{bdi}{BDI}{Belief-Desire-Intention}

\newacronym{prs}{PRS}{Procedural Reasoning System}

\newacronym{are}{ARE}{Autonomous Reasoning Engine}

\newacronym{dmars}{dMARS}{distributed Multi-Agent Reasoning System}

\newacronym{fsm}{FSM}{Finite-State Machine}

\newacronym{pfsm}{PFSM}{Probabilistic Finite-State Machine}

\newacronym{apfsm}{APFSM}{\textit{Autonomous} Probabilistic Finite-State Machine}

\newacronym{pha}{PHA}{Probabilistic Hybrid Automata}

\newacronym{fsa}{FSA}{Finite-State Automata}

\newacronym{ta}{TA}{Timed Automata}

\newacronym{gspn}{GSPN}{Generalised Stochastic Petri Net}

\newacronym{mopn}{MOPN}{Marked Ordinary Petri Net}

\newacronym{ba}{BA}{B\"{u}chi Automata}

\newacronym{dtmc}{DTMC}{Discrete-Time Markov Chain}

\newacronym{pars}{PARS}{Process Algebra for Robot Schemas}

\newacronym{fsp}{FSP}{Finite State Processes}

\newacronym{piadl}{$\pi$ADL}{$\pi$ calculus combined with ADL}

\newacronym{csp}{CSP}{Communicating Sequential Processes}

\newacronym{ccs}{CCS}{Calculus of Communicating Systems}

\newacronym{wsccs}{WSCCS}{Weighted Synchronous CCS}

\newacronym{csp|b}{CSP$\|$B}{an integrated formalism combining CSP and B}

\newacronym{fdr}{FDR}{Failures-Divergences Refinement}

\newacronym{jpf}{JPF}{Java PathFinder}

\newacronym{ajpf}{AJPF}{Agent Java PathFinder}

\newacronym{mcmas}{MCMAS}{Model Checker for Multi-Agent Systems}

\newacronym{ltlmop}{LTLMoP}{Linear Temporal Logic MissiOn Planning}

\newacronym{adl}{ADL}{Architecture Description Language}

\newacronym{aadl}{AADL}{Architecture Analysis and Design Language}

\newacronym{lts}{LTS}{Labelled-Transition System}

\newacronym{uml}{UML}{Unified Modelling Language}

\newacronym{sysml}{SySML}{Systems Modelling Language}

\newacronym{robotml}{RobotML}{Robot Modelling Language}

\newacronym{dsl}{DSL}{Domain Specific Language}

\newacronym{sct}{SCT}{Supervisory Control Theory}

\newacronym{ide}{IDE}{Integrated Development Environment}

\newacronym{mde}{MDE}{Model-Driven Engineering}

\newacronym{ai}{AI}{Artificial Intelligence}

\newacronym{fdir}{FDIR}{Fault Detection, Isolation and Recovery}

\newacronym{ifm}{iFM}{Integrated Formal Methods}

\newacronym{ros}{ROS}{Robot Operating System}

\newacronym{dl}{\text{\upshape\textsf{d{\kern-0.05em}L}}}{differential dynamic logic}

\newacronym{vm}{VM}{Virtual Machine}

\newacronym{bip}{BIP}{Behaviour Interaction Priority}

\newacronym{fol}{FOL}{First-Order Logic}

\newacronym{owl}{OWL}{Web Ontology Language}

\newacronym{ria}{RIA}{Restore Invariant Approach}

\newacronym{ocl}{OCL}{Object Constraint Language}

\newacronym{sil}{SIL}{Safety Integrity Level}

\newacronym{rcl}{RCL}{ROS Contract Language}
\usepackage{paralist}

\usepackage[hidelinks]{hyperref}

\usepackage{zed-csp}
\newcommand{\cspseq}{\,;}

\usepackage{tabularx}

\usepackage[disable, colorinlistoftodos]{todonotes}

\begin{document}
\title{Formal Verification of a Map Merging Protocol in the Multi-Agent Programming Contest\thanks{Work supported by UK Research and Innovation, and EPSRC Hubs for ``Robotics and AI in Hazardous Environments'': EP/R026092 (FAIR-SPACE), EP/R026173 (ORCA), and EP/R026084 (RAIN). The formal specification work was mostly done while the first author was employed by the University of Manchester, UK.}}

%
%
\author{Matt Luckcuck\orcidicon{0000-0002-6444-9312} \\ \small{Department of Computer Science, Maynooth University, Ireland} \and 
Rafael C. Cardoso\orcidicon{0000-0001-6666-6954} \\ \small{Department of Computer Science, The University of Manchester, United Kingdom}}

\maketitle              
\begin{abstract}
\noindent Communication is a critical part of enabling multi-agent systems to cooperate. This means that applying formal methods to protocols governing communication within multi-agent systems provides useful confidence in its reliability. In this paper, we describe the formal verification of a complex communication protocol that coordinates agents merging maps of their environment. The protocol was used by the LFC team in the 2019 edition of the Multi-Agent Programming Contest (MAPC). Our specification of the protocol is written in \gls{csp}, which is a well-suited approach to specifying agent communication protocols due to its focus on concurrent communicating systems. We validate the specification's behaviour using scenarios where the correct behaviour is known, and verify that eventually all the maps have merged.

\end{abstract}
\glsresetall
\section{Introduction}
\setcounter{footnote}{0}

The Multi-Agent Programming Contest\footnote{\url{https://multiagentcontest.org/}.} (MAPC) is an annual challenge to foster the development and research in multi-agent programming. Every couple of years a new challenge scenario is proposed, otherwise, some additions and extensions are made to make the scenario from the previous year more challenging.

The 2019 edition of MAPC~\cite{MAPC19} introduced the Agents Assemble scenario, where two teams of multiple agents compete to assemble complex block structures. Agents have incomplete information about their grid map environment. They are only able to perceive what is inside their limited range of vision. Therefore, building a map of the team's environment must be done individually at the start; each agent believes its starting position is (0,0), and the agents merge their maps when they meet, adjusting the coordinates accordingly.

This paper describes the formal specification and verification of the map merge protocol that was used by the winning team from MAPC 2019, the Liverpool Formidable Constructors (LFC)~\cite{Cardoso20d}. One of the major challenges in the MAPC is making sure all critical parts of the team work reliably. Without a coherent map, the agents cannot coordinate to assemble the block structures, so the map merge protocol is critical to the team's mission. The complexity of the different challenges in the scenario, as well as the presence of another interfering team, can cause unforeseen problems. Before MAPC 2019 the LFC team had limited time to perform tests to validate the code, which due to its complexity meant that it was very hard to efficiently prevent or detect bugs. 

In this paper we build a formal specification of the map merge protocol from its implementation and previous description in~\cite{Cardoso20d}, and formally verify the specification to provide evidence of the protocol's reliability. Our specification is written in the process algebra \gls{csp}~\cite{Hoare1978}, which is designed for specifying concurrent communicating systems. We view each agent in the system as a process, which is communicating with the other agents (processes) to achieve the system's overall behaviour. 

To verify properties about our specification we use model checking, which can automatically and exhaustive check the state space for a formal model for satisfaction of a given property. If a property is violated, a model checker usually gives a counterexample, which can aid debugging. In \gls{csp}, model checking uses the idea of \textit{refinement}. If we have two specifications $P$ and $Q$, then $P$ is refined by $Q$ ($ P \sqsubseteq Q$) if every behaviour of $Q$ is also a behaviour of $P$. This can be thought of as $Q$ implementing $P$, like a software component implementing an interface. We use the \gls{csp} model checker \gls{fdr}~\cite{GibsonRobinson2014} to show that the system behaves according to some required properties. This can be thought of as checking that the system \textit{correctly} implements an interface. 

The work presented in this paper is motivated in two directions. First, the verification provides extra confidence that the protocol works. The protocol was difficult to test because of the dynamic environment and the amount of agent communication, but model checking is a useful approach to finding corner cases. Second, the MAPC provides an interesting example application to explore the utility of \gls{csp} for modelling this kind of problem. This is of lesser importance than the first motivation, but useful nonetheless. 

The rest of this paper is organised as follows. A brief background on JaCaMo (the language the agent system is developed in) and \gls{csp} is presented in the next section, Sect.~\ref{sec:back}. Section~\ref{sec:main} describes how we used \gls{csp} to specify and verify the map merge protocol used by the LFC team in the MAPC 2019. It contains a detailed description of how the protocol works (Sect.~\ref{sec:mergeProtocol}), the \gls{csp} specification of the protocol (Sect.~\ref{sec:cspModel}), and how the specification was validated and verified (Sect.~\ref{sec:verification}). The related work is discussed in Sect.~\ref{sec:rw}, with a variety of similar approaches that have been applied to the specification and verification of multi-agent systems. Finally, Sect.~\ref{sec:conc} presents our concluding remarks.

\section{Background}
\label{sec:back}

The LFC team uses the JaCaMo multi-agent programming platform to develop their agents for the MAPC 2019. In this section we briefly explain JaCaMo and highlight the relevant parts that were used in the map merging protocol (Sect.~\ref{sec:jacamo}). We also give an overview of \gls{csp} and the notation used throughout the paper, and introduce model checking (Sect.~\ref{sec:cspIntro}).

\subsection{JaCaMo}
\label{sec:jacamo}

JaCaMo\footnote{\url{http://jacamo.sourceforge.net/}.}~\cite{Boissier11,boissier2020multi} is a multi-agent development platform that combines three dimensions that are often found in agent systems (agent, environment, and organisation), and provides first-class abstractions that enable a developer to program these dimensions in unison. JaCaMo is a combination of three different technologies that were developed separately and then linked together: the Jason~\cite{Bordini07} \gls{bdi}~\cite{rao:95b} agent programming language for the agent dimension, CArtAgO~\cite{Ricci09} for programming environments using artefacts, and Moise~\cite{Hubner07} for the specification of organisation of agents. An additional first-class abstraction has been developed that provides an interaction dimension for JaCaMo in~\cite{Interaction}, but this is not yet fully integrated.

The merge protocol as implemented by LFC~\cite{Cardoso20d} is comprised of message passing between agents and updating information. The agent communication is implemented solely in Jason, while some of the information updates are done in a shared artefact (called the \emph{TeamArtifact}). In this paper we focus on the communication, which is the critical part of the protocol. 

In Jason, communication between agents is based on speech-act theory, where agents send a performative such as \emph{tell} (sends a belief to an agent, causing a belief addition event) or \emph{achieve} (sends a goal to an agent, causing a goal addition event). The formal semantics of speech-act theory for Jason can be found in~\cite{Vieira07}.

\subsection{CSP}
\label{sec:cspIntro}

\gls{csp} is a formal language for specifying the behaviour of concurrent communicating systems. We use the \gls{fdr}~\cite{GibsonRobinson2014} model checker to both manually and automatically check specifications. Manual checks use \gls{fdr}'s Probe tool, which enables a user to step through the system's behaviour. Automatic checks are written as assertions.

A \gls{csp} specification is built from (optionally parameterised) processes. A process describes behaviour as a sequence of \textit{events}; for example, $a \then b \then \Skip$ is the process where the events $a$ and $b$ happen sequentially, followed by $\Skip$ which is the terminating process. An event\footnote{Note that events in CSP are different from Jason BDI events, the former are communication events while the latter are plan triggering events.} is a communication on a \textit{channel}. Channels enable message-passing between processes, but a process can perform an event (communicate the event on the channel) even if there is no other process to receive the event. Where two processes agree to perform a set of events in parallel (\textit{synchronise} on a set of events) both processes must perform the event(s) synchronously.

By convention \gls{csp} process names are written in upper-case, and channels or events in lower-case. A \gls{csp} process is often composed of several `subprocesses', which is the term we use to refer to other processes called by a process. Here, a subprocess helps to structure the specification and encapsulate behaviour, similarly to an object and its methods. We adopt the convention of using a double underscore to separate a process name from the `main' process to which it belongs (e.g., $MAIN\_PROCESS\_\_SUBPROCESS$). Below, we describe the \gls{csp} operators used in this paper, which are also summarised in Table~\ref{tab:cspOperators}.

\begin{table}[t]
\centering
\begin{tabularx}{\textwidth}{ l | l | X  }
    \hline
Action			& Syntax 				& Description \\
\hline \hline
Skip				& $\Skip$ 				& The terminating process \\
\hline
Simple Prefix	&  $a \then \Skip$		& Simple synchronisation on $a$ with no data, followed by $\Skip$ \\
\hline
Input Event 		& $a?in$					& Synchronisation that binds a the input value to $in$ \\
\hline
Output Event		& $b!out$				& Synchronisation outputting the value of $out$\\
\hline
Parameter Event	& $c.value$				& Synchronisation matching the given $value$ \\
\hline
Sequence			& $P~\cspseq{}Q$			& Executes processes $P$ then $Q$ in sequence\\
\hline
External Choice & $P \extchoice Q$		& Offers a choice between two processes $P$ and $Q$\\
\hline
Replicated External Choice & $\Extchoice x : Set @ P(x)$ & Offers an external choice of the process $P(x)$ with every value $x$ in the set $Set$ \\
\hline 
Parallelism 		& $P \parallel[chan] Q$ 	& $P$ and $Q$ run in parallel, synchronising on the channels in $chan$ \\
\hline
Parallelism		& $P \parallel[pChan][qChan] Q$ & $P$ and $Q$ run in parallel, synchronising on the channels common to the sets $pChan$ and $qChan$ \\
\hline
Interleaving 	& $P \interleave Q$ 		& $P$ and $Q$ run in parallel with no synchronisation \\
\hline
Replicated Interleaving & $\Interleave x: Set @ P(x) $ & Interleaves a copy of the process $P(x)$ for every value $x$ in the set $Set$ \\
\hline \hline
\end{tabularx}
\caption{Summary of \gls{csp} operators used in this paper \label{tab:cspOperators}}
\end{table}

Channels may declare typed parameters. If a channel is untyped, then we get `simple' events like $a$ and $b$ from the above example; the events of a typed channel must contain parameters matching those types. 
For example, if channel $c$ takes one integer parameter, then an event might be $c~.~42$. 
Parameters may be inputs ($c?in$), outputs ($c!out$), or match a given value ($c.value$). Inputs can be restricted ($c?p:set$) to only accept a parameter ($p$) that is the given $set$ (here, a set of integers). 
Processes can occur in sequence; for example, $P~\cspseq{}Q$ describes a process where process $P$ runs, then process $Q$. 

A choice of processes can be offered; for example $P \extchoice Q$ offers the choice of either $P$ or $Q$, once one process is picked the other becomes unavailable. Processes can also run in parallel.
\gls{csp} provides three parallel operators; in $P \parallel[ chan ] Q$, processes $P$ and $Q$ run in parallel, and agree to communicate on the channels in the set $chan$; in $P \parallel[pChan][qChan] Q$, processes $P$ and $Q$ run in parallel, and agree to communicate on the channels common to the $pChan$ and $qChan$ sets; and in $P \interleave Q$, the processes $P$ and $Q$ run at the same time with no synchronisation. 

\gls{csp} does not have variables, so if a specification needs a variable that will be accessed by several processes, a `state process' is often used. This is where a process is used to store, and control access to, some values. The values are stored as process parameters and channels are provided to get and set the values. Other processes communicate with the state process using these get and set channels. While this requires more channels and internal communication, it can lead to cleaner communication between the processes that need to use the variable(s).

\section{Specification and Verification of the Map Merge Protocol}
\label{sec:main}

The map merge protocol played a major role in LFC's victory\footnote{Source code of the team is available at (the main plans for the map merge protocol are located in ``src/agt/strategy/identification.asl''): \url{https://github.com/autonomy-and-verification-uol/mapc2019-liv}.}~\cite{Cardoso20d} in MAPC 2019. It overcomes one of the main challenges that has to be solved before trying to assemble structures and deliver tasks. Even though LFC won, there were many failures during the matches due to limited testing and no formal verification prior to the contest. Although the origin of most of the failures is still unknown, we aim to provide confidence about the reliability of the map merge protocol.

In this paper we specify and verify the map merge protocol as used by LFC~\cite{Cardoso20d}. Section~\ref{sec:mergeProtocol} describes the protocol, in particular the communication between the agents. Then, Sect.~\ref{sec:cspModel} presents our \gls{csp} specification of the protocol and describes how it models two agents merging their maps. Finally, Sect.~\ref{sec:verification} discusses the validation and verification of the specification using the \gls{fdr} model checker.

\subsection{Map Merge Protocol}
\label{sec:mergeProtocol}

\begin{figure}[ht]
    \centering
    \includegraphics[width=\linewidth]{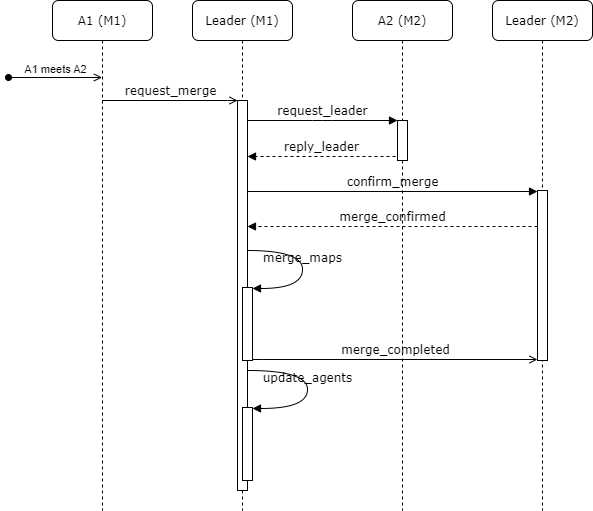}
    \caption{UML Sequence diagram for the map merge protocol. \texttt{A1} represents Agent1 with a map \texttt{M1}, \texttt{A2} is Agent2 with a map \texttt{M2}. The solid arrow heads represent synchronous messages, open arrow heads asynchronous messages, dashed lines represent reply messages, and rectangles represent processes~\cite{Cardoso20d}.}
    \label{fig:merge}
\end{figure}

The communication in the map merge protocol consists of message passing between a group of agents, triggering plans that reason about the message received and send the required reply if applicable. 
Figure~\ref{fig:merge} shows an overview of the protocol. At the start of the simulation, every agent has its own map (referred to as being the \emph{leader} of its own map, or being a \textit{map leader}). 
As the simulation progresses, agents meet each other and merge their maps. Each map leader coordinates a map for itself and any other agents whose maps it has merged with. This means that there can be a minimum of two and a maximum of four agents directly involved in one instance of the merge protocol. For example, only two agents will be involved if both agents are the leaders of their own map.

Each agent has a name, or ID; in Fig.~\ref{fig:merge} we use \texttt{A1} and \texttt{A2}. Because of the (intentionally challenging) communications restrictions of the MAPC, when agents meet they do not know each other's ID, so they cannot directly communicate. Therefore, before agents can exchange useful information about their maps, some kind of \emph{identification} process is needed. In LFC's strategy for identification, when one agent sees another it sends a broadcast to all the agents in its team requesting information about what they can see around them. Upon receipt of all of the replies, the agent that sent the request will compare the replies to what it can see, to try to identify the agents that it has met. The specifics of LFC's identification strategy can be found in~\cite{Cardoso20d}.

The merge protocol starts as part of the identification strategy, after one agent successfully identifies another. The identification process is reflexive but asynchronous. For example, if \texttt{A1} identifies \texttt{A2}, then \texttt{A2} will (eventually) identify \texttt{A1}; but they each perform the identification process separately, so they might not both identify each other at the same time. In this example each agent will then start its own merge process, but as we will see later on only one merge will go through. In Fig.~\ref{fig:merge} the protocol is seen from \texttt{A1}'s perspective, but it works exactly the same for all agents.

Once the map merge protocol starts, the agent that is requesting the merge (the \textit{requesting agent}) sends a message to its map leader (now, the \textit{requesting map leader}) containing a list of agents that it wants to merge with, that is, all agents that it has successfully identified. Requests are dealt with one at a time, so each request constitutes a new instantiation of the protocol. As previously mentioned, messages in Jason trigger events which will enable plans that match the triggering event to be viable for selection\footnote{A plan still has to succeed its context check (i.e., meet its preconditions) before being selected for execution.}. The requesting agent's map leader will start the plan that handles the request to merge, when it receives the \texttt{request\_merge} message; if the requesting agent is its own map leader it will simply trigger the appropriate plan internally.

To proceed with the merge the requesting map leader must get the name of the \emph{other} agent's map leader (for example, in Fig.~\ref{fig:merge} agent \texttt{A1} wants to merge maps with \texttt{A2}). After receiving the name of the other map leader, the requesting map leader will use a priority order among both map leaders to determine if the merge will continue. This is necessary because the other agent may also have started the map merge process , however, only one of these merges can proceed. The priority is determined using the number in the agent's name, the agent with the 
lowest number has priority. For example agent \texttt{A1} has priority over \texttt{A2}.

After the priority check, if the merge is continuing then the requesting map leader sends a message to the other map leader asking it to confirm the merge. This is necessary because it is possible that the other map leader is already in the middle of a merge, which could result in it losing its position as a map leader. If this happens and the original merge continues, then the map information from this merge will be wrong. Thus, if the other map leader is no longer the leader of that map it will send a reply cancelling the merge. Another attempt to merge these two maps can be made in the next step.

The plans for both leaders are atomic, so concurrent intentions are not in effect (normally in Jason agents alternate between their intentions). This means that leaders are not able to start multiple merge processes simultaneously, nor can they enter a deadlock waiting for information indefinitely (we assume that eventually all agents reply).

If the merge is confirmed, then the map information (such as the coordinates of points of interest in the MAPC scenario) is updated. Finally, a message is sent to the other map leader letting it know that the merge has been completed, and thus releasing the lock from the atomic plan that it was in. A final update is made to the list of agents that are part of the new merged map, which is sent to each agent in the list.

In summary, the main goal and requirements of the map merge protocol are: 
\begin{itemize}
    \item[GOAL:] If agent $A1$ merges its map with agent $A2$, $A1$'s map leader will be the map leader for $A1$, $A2$, and any agents that either of them shared their map with before the merge.
    \item[REQ1:] The map leader of $A1$ has priority over the map leader of $A2$, otherwise $A1$ should cancel the merge.
    \item[REQ2:] If $A2$'s map leader loses control of its map by the time it processes the request to merge from $A1$'s map leader, it will cancel the merge.
\end{itemize}
In the next section we describe how we model the map merge protocol, while Sect.~\ref{sec:verification} describes how we verify that the protocol preserves these properties.

\subsection{CSP Specification}
\label{sec:cspModel}

Our \gls{csp} specification\footnote{The \gls{csp} files are available at: \url{https://doi.org/10.5281/zenodo.4624507}.} was built by one person, part-time over the course of about 12 months. The specification was built by manually interpreting and translating the English-language description of the protocol~\cite{Cardoso20d} and the Jason implementation. 
The specification is $\sim$440 lines in total, though this includes comments. Despite the small number of lines, the specification contains 1,597,190 states and 6,334,936 transitions\footnote{Reported states from a check for freedom from non-determinism using \gls{fdr} 4.2.7.} . 

\begin{sloppypar}
In the specification of the map merge protocol, each agent is modelled by an $AGENT$ process, if the agent is a map leader then it will also be represented by a $MAP\_LEADER$ process. As mentioned in Sect.~\ref{sec:mergeProtocol}, at the start of a match each agent is also its own map leader, so each agent will begin as a cooperating pair of $AGENT$ and $MAP\_LEADER$ processes 
The `main' processes, $AGENT$ and $MAP\_LEADER$, are decomposed into subprocesses that structure the specification and encapsulate behaviour. We remind the reader that we use a double underscore to separate a subprocess name from the `main' process to which it belongs. For example, the subprocess named
$MAP\_LEADER\_\_REQUEST\_MERGE$ is the $REQUEST\_MERGE$ processes which belongs to the $MAP\_LEADER$ process.
\end{sloppypar}

\begin{sloppypar}
Our specification only uses three agents. This was a conscious choice to keep the specification's state space small, while still enabling us to check that a pair of agents can merge their maps in the presence of an interfering agent. 
We define $AgentName$, the set of all the agent IDs as:
\smallskip
\centerline{$datatype~AgentName = A1 | A2 | A3$}
\smallskip
\noindent The top-level process of our specification is a parallel composition ($ \parallel[chan]$) of all the $AGENT$ and $MAP\_LEADER$ processes. This takes the form:
\smallskip
\centerline{$LFC = AGENTS \parallel[interface] MAP\_LEADERS $}
\smallskip

\noindent where $AGENTS$ is an interleaving ($\interleave$) of an $AGENT$ process for each ID in $AgentName$; and $MAP\_LEADERS$ is a parallel composition of a $MAP\_LEADER$ process for each ID in $AgentName$, so the map leaders can communicate. In this top-level process, the $AGENT$ and $MAP\_LEADER$ processes agree to synchronise (cooperate) on any events in the set of events $interface$.
\end{sloppypar}

$MAP\_LEADERS$ composes the $MAP\_LEADER$ processes in a way that allows them to synchronise on some events, because they communicate to control the map merge. But $AGENTS$ simply interleaves the $AGENT$ processes because they do not need to communicate for the merge protocol. Other behaviours require agents to communicate, but this is abstracted away in our specification.

\begin{sloppypar}
The IDs in $AgentName$ (for example $A1$) are used to synchronise communications between a $MAP\_LEADER$ and the $AGENT$  processes that it coordinates, and between it and other $MAP\_LEADER$ processes. The messages in Fig.~\ref{fig:merge} are represented in our specification by \gls{csp} channels. The processes in the specification also make use of other internal channels to describe the required behaviour.
\end{sloppypar}

As mentioned in Sect.~\ref{sec:mergeProtocol}, a map merge happens between two agents and their respective map leaders (note that an agent may be its own map leader). Unlike the description in Sect.~\ref{sec:mergeProtocol}, there will \textit{always} be four processes involved in a map merge in our specification: two $AGENT$ processes and two $MAP\_LEADER$ processes. This is because we model the behaviours common to all agents separately from the behaviours specific to a map leader. As in Sect.~\ref{sec:mergeProtocol}, we refer to the agent that requests the merge and its map leader as the `requesting agent' and the `requesting map leader', respectively. 

We use the example from Fig.~\ref{fig:merge} to describe how our model captures the scenario. We have split this into three phases: 
\begin{compactenum}
    \item \textbf{Requesting Merge and Leader:} initial communication to obtain information that will be used during the map merge and used to check the viability of the merge;
    \item \textbf{Confirming Merge:} use the information obtained in the previous phase to determine if the merge can proceed or if it should be cancelled;
    \item \textbf{Merge and Update:} perform the map merge and update all agents involved.
\end{compactenum}
As in Fig.~\ref{fig:merge}, agent $A1$ is requesting that agent $A2$ merges maps with it, and both agents are their own map leader. Because we do not include the maps themselves in our specification, we do not use the map IDs ($M1$ and $M2$).

\subsubsection{Phase 1: Requesting Merge and Leader} \label{sec:requestingMergeandLeader}~\\

\noindent~In the first phase of the protocol, an agent sends a request to its map leader to merge with one or more other agents, then the map leader requests the name (ID) of the mad leader of each of the agents it has been requested to merge with. When one of the other agents replies, the requesting map leader begins negotiating the map merge with the other map leader. 

In our example, $A1$'s $AGENT$ process sends a message to its map leader (which is itself) requesting that it merges with $A2$. This is represented by an event on the channel $request\_merge$, which is sent from the $AGENT$($A1$) process to the $MAP\_LEADER$($A1$) process and triggers the $MAP\_LEADER$ to start its map merging process. In the \textit{requesting} $MAP\_LEADER$, this is handled by the $MAP\_LEADER\_\_REQUEST\_MERGE$ process, part of which is shown in Fig.~\ref{fig:requestMerge}

\begin{figure}[t]
    \centering
\begin{syntax}
MAP\_LEADER\_\_REQUEST\_MERGE(Me, AgentSet) = \\
\t1 MAP\_LEADER\_\_HANDLE\_REQUEST\_MERGE(Me, AgentSet) \\
\t1 \parallel[ \{| beginMerge|\}] \\
\t1 MAP\_LEADER\_\_BEGIN\_MERGE(Me, AgentSet)
\end{syntax}

\caption{Excerpt from $MAP\_LEADER\_\_REQUEST\_MERGE$ showing the parallel composition of  $MAP\_LEADER\_\_HANDLE\_REQUEST\_MERGE$, which handles requests to merge; and 
$MAP\_LEADER\_\_BEGIN\_MERGE$, which allows the map leader to begin a merge. The processes synchronise on the $beginMerge$ event. $Me$ is the ID of the $MAP\_LEADER$ process, and $AgentSet$ is the set of agents this map leader is coordinating. }
\label{fig:requestMerge}
\end{figure}

The $MAP\_LEADER\_\_HANDLE\_REQUEST\_MERGE$ process listens for $request\_merge$ events from any agent it currently coordinates (any agent in its $AgentSet$). The $request\_merge$ event contains a parameter $mergeSet$, which is the set of agents that the map leader should try to merge with. In our example, $merge\_set$ only contains $A2$, but it can contain the IDs of any agents that are not in the $AgentSet$ of the requesting map leader (as explained in Sect.~\ref{sec:mergeProtocol}). 

When a $request\_merge$ event is received, the $beginMerge$ event is used to start the merge. This is an event introduced purely for our specification. The $MAP\_LEADER\_\_BEGIN\_MERGE$ process sends $request\_leader$ events to each agent in the $mergeSet$ and waits for a $reply\_leader$ event from one of these agents, acting on the first of these events to arrive. The $reply\_leader$ event contains a parameter that is the ID of the agent's map leader. In our example, $A2$ sends the reply that its map leader is $A2$. 

The $MAP\_LEADER\_\_BEGIN\_MERGE$ process, triggered by $A1$ receiving the $reply\_leader$ event from $A2$, checks if the map leader for $A1$ has priority over the map leader for $A2$.  As mentioned in Sect.~\ref{sec:mergeProtocol}, the agent with the lowest ID number takes priority, so $A1$ has priority over $A2$, which has priority over $A3$.  If the requesting map leader does not have priority, then the map merge attempt ends here. In our example, the map leader is the same as the requesting agent, $A1$, and $A1$ does have priority over $A2$ (the other map leader), so it moves on to confirming the merge with the other $MAP\_LEADER$ process.

\subsubsection{Phase 2: Confirming Merge} \label{sec:confirmMerge} ~\\

\noindent The requesting map leader asks the other map leader to confirm that the merge can proceed. As explained in Sect.~\ref{sec:mergeProtocol}, this allows a map leader to cancel a merge request if it has completed a (concurrent) merge with a different map leader. In our example, the requesting map leader is $A1$ and the other map leader is $A2$.

\begin{sloppypar}
The requesting $MAP\_LEADER$ ($A1$) handles this phase using the process $MAP\_LEADER\_\_CONFIRMING\_MERGE$ (excerpt in Fig.~\ref{fig:confirmingMerge}). The first event in this process, $confirm\_merge.Me.OtherMapLeader$, is a request from map leader $A1$ (here, the value of $Me$) to map leader $A2$ (here, the value of $OtherMapLeader$), to confirm that the merge can go ahead. As we can see from Fig.~\ref{fig:confirmingMerge}, the other map leader ($A2$) can reply with either $merge\_cancelled$ or $merge\_confirmed$.
\end{sloppypar}

\begin{figure}[t]
    \centering
\begin{syntax}
MAP\_LEADER\_\_CONFIRMING\_MERGE \\\t1(Me, RequestingAgent, OtherAgent, OtherMapLeader, AgentSet) = \\
 confirm\_merge.Me.OtherMapLeader \then \\
(\\
\t1        merge\_cancelled.Me.OtherMapLeader \then \\
\t1         remove\_reasoning\_about.RequestingAgent.OtherAgent \then \Skip\\
)\\
     \extchoice \\
(\\
\t1        merge\_confirmed.Me.OtherMapLeader?otherAgentSet \then \\
\t1        MAP\_LEADER\_\_MERGE\_MAPS\\\t2(Me, AgentSet, OtherMapLeader, otherAgentSet) \cspseq{} \\
\t1       MAP\_LEADER\_\_UPDATE\_AGENTS\\\t2(Me, AgentSet, OtherMapLeader, otherAgentSet) \\
)
\end{syntax}

\caption{Excerpt from $MAP\_LEADER\_\_CONFIRMING\_MERGE$, which handles the requesting map leader confirming the merge with the other map leader. The parameter $Me$ is the ID of the $MAP\_LEADER$ process, $RequestingAgent$ is the ID of the agent that requested the merge, $OtherAgent$ is the agent the $RequestingAgent$ wanted to merge with, $OtherMapLeader$ is the ID of the other map leader (the leader of $OtherAgent$'s map), and the $AgentSet$ is the set of agents that this map leader is coordinating. }
\label{fig:confirmingMerge}
\end{figure}

\begin{sloppypar}
When the map leader $A2$ receives the $confirm\_merge$ event, it `considers' the eligibility of the merge. If it is no longer a map leader, it replies $merge\_cancelled$, and the merge process will terminate after replying to any pending $confirm\_merge$ events. The $merge\_confirmed$ event signals that the merge can continue. Either of these events passes control back to the requesting map leader ($A1$), which will: $SKIP$ (terminate) if the reply was $merge\_cancelled$; or move on with the merge and update step if the reply was $merge\_confirmed$. 
\end{sloppypar}

\subsubsection{Phase 3: Merge and Update} \label{sec:mergeAndUpdate} ~\\

\noindent This phase is split into two stages: merging the maps, and updating the agents. First, the requesting map leader uses the $MAP\_LEADER\_\_MERGE\_MAPS$ process, shown in Fig.~\ref{fig:mergeMaps}, to merge the maps. The process is triggered by the $merge\_confirmed$ event, as shown in Fig.~\ref{fig:confirmingMerge}. 

Since the map is not captured in our specification, merging the maps is abstracted to the event $merge\_maps$. The requesting $MAP\_LEADER$ updates the $AgentSet$ (the set of agents it coordinates). It then tells the other map leader that the merge is completed, using the $merge\_completed$ event, which also sends the union of the two agent sets.

\begin{figure}[t]
    \centering
\begin{syntax}
MAP\_LEADER\_\_MERGE\_MAPS(Me, MyAgentSet, otherMap, OtherAgentSet) =\\
\t1      merge\_maps.Me.otherMap \then \\
\t1 update\_agentSet.Me!OtherAgentSet \then  \\
\t1      merge\_completed.Me.otherMap.union(MyAgentSet, OtherAgentSet) \then \\
\t1  SKIP
\end{syntax}
    \caption{The $MAP\_LEADER\_\_MERGE\_MAPS$, which controls the merging of maps between two map leaders (abstracted to the $merge\_maps$ event. $Me$ is the ID of the $MAP\_LEADER$ process and $MyAgentSet$ is its agent set. Similarly, $otherMap$ is the ID of the other $MAP\_LEADER$ process and $OtherAgentSet$ is its agent set. \label{fig:mergeMaps}}
\end{figure}

\begin{sloppypar}
After $merge\_completed$, control returns to the $MAP\_LEADER\_\_CONFIRMING\_MERGE$ process (Fig.~\ref{fig:confirmingMerge}), which calls the $MAP\_LEADER\_\_UPDATE\_AGENTS$ process to update all the agents in its new $AgentSet$. This involves a sequence of communications between the requesting map leader and each $AGENT$ in the new $AgentSet$. 
\end{sloppypar}

This phase closely corresponds to the description in \cite{Cardoso20d}, summarised as follows:
\begin{compactenum}
\item \textbf{Build new list of identified agents:} in our specification, this is simply the union of the requesting and other map leader's $AgentSet$s.   
\item \begin{sloppypar} \textbf{Send update to the leader of $M2$ ($A2$):} as shown in Fig.\ref{fig:mergeMaps}, the $merge\_completed.Me.otherMap.union(MyAgentSet, OtherAgentSet)$ event sends the new (merged) $AgentSet$ to the other map leader. \end{sloppypar}
\item \begin{sloppypar} \textbf{Send update to all agents of $M1$ ($A1$):} here a recursive process sends $update\_identified\_same\_group$ to each agent in the agent set of the requesting map leader ($A1$). This event passes the new $AgentSet$ to each of these agents. \end{sloppypar}
\item \textbf{Send update to all agents of $M2$ ($A2$):} here a recursive process sends $update\_identified$ to each agent in the $AgentSet$ of the other map leader. This event passes the new agent set to each of these agents and is also used to update each agent's map leader to the requesting map leader, $A1$.
\end{compactenum} \medskip

\begin{sloppypar} \noindent  After the updates are completed, the requesting map leader recurses back to the $MAP\_LEADER\_\_REQUEST\_MERGE$ process (Fig.~\ref{fig:requestMerge}) ready to begin another merge. If there are no more agents to merge with from this request, it waits for the next merge request. The other map leader process no longer represents an agent that is a map leader, so it will only reply $merge\_cancelled$ if it is asked to merge. This handles requests to merge that may already be in progress. 
\end{sloppypar}

\subsection{Specification Validation and Verification}
\label{sec:verification}

After specifying the map merge protocol, we validate that it performs the protocol's required behaviour and then verify that all the maps are eventually merged. The validation step is used to check that the specification conforms with the protocol's implementation. The verification step is used to check that the specification is correct.
For both of these steps, we use the assertions and in-built tools of the \gls{csp} model checker, \gls{fdr}. The assertions are described in Table~\ref{tab:assertions} and the time that FDR took to check the assertions is summarised in Table~\ref{tab:times}.

\begin{table}[t]
\centering
\begin{tabularx}{\textwidth}{ p{5em} | l | X  }
\textbf{Name} & \textbf{Type} & \textbf{Description} \\
\hline \hline
Scenario 1 (REQ1) & has trace & $A1$ merging with $A2$, $A1$ has priority, $A2$ merges into $A1$ \\
\hline

Scenario 2 (REQ2) & has trace & $A1$ merging with $A2$, but $A2$ cancels the merge \\
\hline

Scenario 3 (REQ1) & has trace &  $A2$ merging with $A1$, denied because $A2$ does not have priority \\
\hline 

Scenario 4a (REQ1) & has trace & $A2$ requests a merge with $A3$, then $A1$ requests a merge with $A3$. $A1$ merges with $A3$ first, then $A3$ replies that its Map Leader is now $A1$. $A2$ now tries to merge with $A1$, which is denied because $A2$ does not have priority \\
\hline

Scenario 4b (REQ2) & has trace & $A2$ requests a merge with $A3$, then $A1$ requests a merge with $A3$. $A1$ merges with $A3$ first, then $A3$ replies that its Map Leader is still $A3$. $A2$ tries to merge with $A3$, which is cancelled because $A3$ is not a Map Leader any more \\

\hline
Scenario 5 (REQ1) & has trace & $A1$ merges with $A2$, then $A3$ tries to merge with $A2$, which replies that its Map Leader is now $A1$. $A2$ tries to merge with $A1$, which is denied because $A3$ does not have priority \\

\hline
$done$ Reachable (GOAL)  & refinement & Can the $LFC$ process reach the state where any of the agents can call $done$ (showing that it is coordinating all the agents). \\

\hline \hline
\end{tabularx}
\caption{Summary of the verification (Scenarios 1--5) and validation ($done$ reachable) assertions applied to the map merge protocol. The requirement, or goal, that each assertion covers is presented in brackets. \label{tab:assertions}}
\end{table}

\subsubsection*{Validation}
For the validation step, first we used \gls{fdr}'s Probe tool to manually step through the model one event at a time. This was useful when debugging the specification, especially after adding or updating behaviour. For more substantial verification, we also checked how the agents behaviour in six different scenarios (see Table~\ref{tab:assertions}). These scenarios were based on the implementation's behaviour in LFC's matches during MAPC 2019, so the correct behaviour is known. The specification was checked to see that it would perform each of the scenarios correctly, showing that it corresponds to the implemented protocol.

The scenarios were developed alongside the specification, and were useful for checking that it continued to meet the requirements while behaviour was being added. Hence, Scenarios 1 to 3 describe the requirements of a pair of agents; while Scenarios 4a, 4b, and 5 check the requirements with the interference of a third agent; mirroring the specification's development. 
Scenario 1 is the example in Fig.~\ref{fig:merge}, where agent $A1$ meets agent $A2$ and requests they merge maps, $A1$ has priority so $A2$'s map is merged into $A1$'s; and Scenario 2 shows $A2$ cancelling the merge instead.
Scenario 3 is $A2$ trying to merge with $A1$ and not having the priority to do so. 
Scenarios 4a and 4b check the two situations that can occur when an agent stops being a Map Leader after a third agents has started merging with it. 
Finally, Scenario 5 checks the combination of an agent that stops being a Map Leader and an agent that doesn't have priority for a merge.

While the six scenarios are not an exhaustive list, they cover both of the protocol's requirements (Sect.\ref{sec:mergeProtocol}). 
REQ1, that the merge will be denied if the requesting map leader does not have priority, is checked by Scenarios 1 and 3 (for two agents) and Scenarios 4a and 5 (for three agents). REQ2, that an agent will cancel a merge if it loses control if its map, is checked by Scenario 2 (for two agents) and Scenario 4b (for three agents). The GOAL is checked by the $done$ reachable assertion, described below alongside the other verification checks.

Each scenario is described as a trace of the relevant events in the scenario. We used \gls{fdr}'s in-built $[has~trace]$ check, to explore the model's state space to see if it can perform the scenario trace (though this does not show that it will \textit{always} perform the scenario trace). In the assertion check we hide all the events that are not in the scenario trace. This takes the form:\\
\centerline{$assert~LFC \setminus(diff(Events, trace\_events)) :[has~trace]; <trace\_events>$}
\noindent where $LFC$ is the specification's top-level process, $Events$ is the set of all events, and $trace\_events$ is a sequence of events. The $diff()$ function in the hiding operator $\setminus(diff(Events, trace\_events))$ hides only the events not in $trace\_events$. 

The $[has~trace]$ checks act like tests of the specification. They are run automatically by \gls{fdr}, so they are easily repeatable They also provide useful regression tests, which ensures that a change to the specification during this validation and debugging step has not introduced a bug somewhere else. 

\subsubsection*{Verification} 
For verification, we use \gls{fdr}'s in-built assertions to show (by exhaustive model checking) that our specification of the merge protocol is free from divergence and non-determinism. Divergence (livelock) is where the specification performs infinity many internal events, refusing to offer events to the environment. Non-determinism is where the specification may perform several different events, after a given prefix.

\begin{figure}[t]
    \centering
\begin{syntax}
get\_agentSet.Me~?~gotAgentSet \then \\
if~gotAgentSet == AgentName then \\
\t1 done.Me \then terminate.Me \then SKIP \\
else \\
\t1 MAP\_LEADER\_\_REQUEST\_MERGE(Me, gotAgentSet) \\
    \end{syntax}
    \caption{Excerpt from $MAP\_LEADER\_\_REQUEST\_MERGE$, checking that all the maps have merged. This follows on from the excerpt in Fig.~\ref{fig:requestMerge}. }
    \label{fig:doneCheck}
\end{figure}

Finally, we check that the specification can reach a state where all the maps have merged. To get to this state shows that the specification performs the correct behaviour and that it does not deadlock before reaching the `done' state. If the specification reaches this state, it shows that the GOAL and requirements REQ1 and REQ2 are obeyed by the specification.
This check required the addition of the $if\ldots else\ldots$ construct to the $MAP\_LEADER$ process (shown in Fig.~\ref{fig:doneCheck}) which is not part of the map merge protocol. The check happens inside the $MAP\_LEADER\_\_REQUEST\_MERGE$ subprocess, after a merge has been either confirmed or cancelled. The event $get\_agentSet$ retrieves the agent set (the set of agents that this Map Leader is coordinating) from the map leader's internal state process; which is named $gotAgentSet$ here, to avoid a name clash. 

The $gotAgentSet$ is compared to the set of all agent IDs ($AgentName$), using an $if\ldots then\ldots else\ldots$ construction that is not part of \gls{csp} but is available in the input language of \gls{fdr}. If the sets are equal (meaning that this $MAP\_LEADER$ is now coordinating all the agents) then the process synchronises on the $done$ event. In our specification, the $done$ event can happen after a minimum of two successful map merges (agent $A1$ merging with agents $A2$ and $A3$ in either order) but there could be more, depending on the interleaving of events. This means that the $done$ event represents several successful instances of the protocol, each of which must have obeyed the GOAL and requirements REQ1 and REQ2. We can also see in Fig.~\ref{fig:doneCheck} that after $done$, the $MAP\_LEADER$ waits for the $terminate$ event, which tells it to terminate. This is also only part of our specification, not a part of the merge protocol.

To check if the state where a $MAP\_LEADER$ can call $done$ is reachable, we use the following assertion:
\begin{syntax}
assert~LFC\setminus(diff(Events, \{|done|\})) [FD=\\
\t1 \Extchoice agent : AgentName @ done.agent \then SKIP
\end{syntax}
\noindent which checks if the specification ($LFC$) is refined by ($[FD=$) the process that offers the external choice ($\Extchoice$) of any $MAP\_LEADER$ calling $done$. Again, we use $\setminus(diff(Events, \{|done|\}))$ to hide all the events in $LFC$ other than $done$, because it is the only event pertinent to this check. Here, the \textit{replicated} external choice (see Table.~\ref{tab:cspOperators}) offers the $done$ event with each ID $agent$ in the set of all agent IDs, $AgentName$. The particular refinement check used here (in \gls{csp}'s \textit{failures-divergences} model) means that the $LFC$ processes cannot refuse the $done$ event (as this would be a \textit{failure}) and it cannot diverge. As previously mentioned, this shows that $LFC$ does not deadlock before the $done$ event occurs.

\subsubsection*{Discussion}

\begin{table}[t]
\centering
\begin{tabular}{ l | l | l | l  }
\textbf{Name} & \textbf{Compiled (s)} & \textbf{Checked (s)} & \textbf{Total (s)} \\
\hline \hline
Scenario 1 & 0.84 &	0.15 &	0.99 \\

\hline

Scenario 2 & 0.89 &	0.10  &	0.99 \\

\hline

Scenario 3 & 0.89 &	0.10	 & 0.99 \\

\hline 

Scenario 4a & 0.94 &	0.08	 & 1.02 \\

\hline

Scenario 4b & 0.95 & 0.06	& 1.01 \\

\hline
Scenario 5 & 0.86 & 0.10 & 	0.96 \\

\hline

Divergence & 0.71 &	2.41 & 3.12 \\

\hline

Non-Determinism & 6.19 &	 2.69 & 8.88 \\

\hline
$done$ Reachable  & 4.72 & 0.02 & 4.74 \\

\hline \hline
\end{tabular}
\caption{Summary of times (in seconds) taken to check each scenario trace ($has~trace$), divergence, non-determinism, and that the $Done$ event is reachable. The times shown are for a single run using FDR 4.2.7, showing how long it took to compile and check each assertion, the total time is the sum of the compilation and checking time. \label{tab:times}}
\end{table}
\begin{sloppypar}
Table~\ref{tab:times} shows a summary of the times (in seconds) taken to complete the $has~trace$ checks on each scenario trace, the divergence and non-determinism checks, and the check that the $done$ event is reachable. These results are from using FDR 4.2.7, on a PC using Ubuntu 20.04.2, with an Intel Core i5-3470 3.20 GHz × 4 CPU, and 8 GB of RAM. The table reports the compilation time, which is how long it took FDR to build its internal representation of the specification; checking time, which is how long it took FDR to actually check the assertion; and the total time, which is simply the sum of the previous two times. 
\end{sloppypar}

The total times for these verification and validation checks were small enough to not be a barrier to quick re-checking of the properties after updates to the specification. The scenario traces provided quick regression tests, each being checked in $\sim$1s. Even the longest of the three exhaustive checks (non-determinism) was still relatively fast, at only 8.88s in total.

As mentioned in Sect.~\ref{sec:cspModel}, our model only uses three agents, which helps keep the state space of the specification small. To provide a comparison, we added a fourth agent to $AgentName$ (the set of all agent IDs, mentioned in Sect.\ref{sec:cspModel}):\\
\centerline{$datatype AgentName = A1 | A2 | A3 | A4 $}
The model was not \textit{specifically} designed for more than three agents, but it adapts to the number of agent IDs (for example, it runs one $AGENT$ process for each ID in $AgentName$). Then, we rechecked the scenario traces in \gls{fdr}. For three agents they each took $\sim$1s (Table~\ref{tab:times}); for four agents they took between 89s and $\sim$107s longer, an average increase of 9788.94\% (97.20s). We did not compare the times for the exhaustive checks because they used used all the RAM on the test PC, which will artificially increase the checking time. If we add more agents to the model, other elements may need to be altered to reduce the state space. However, this is left for future work.

\section{Related Work}
\label{sec:rw}

A recent survey~\cite{Bakar18} identified that the main validation and verification approaches being applied to agent systems are: model checking, theorem proving, runtime verification, and testing. Testing has been shown to be less effective in the validation and verification of BDI-based agent systems when compared to traditional procedural programs, encouraging the use of formal methods~\cite{Winikoff17}.

Various other approaches for model checking multi-agent systems exist in the literature. MCMAS~\cite{LomuscioR06} and MCK~\cite{MCK2014} are two symbolic model checkers for agent systems, and AJPF~\cite{MCAPL_journal} is a \textit{program} model checker for agents written in the Gwendolen~\cite{dennis08:_gwend} language. Runtime verification has also been used to verify agent interaction protocols specified as trace expressions in~\cite{DBLP:conf/atal/FerrandoAM17}. 

However, these approaches work best when applied top-down, and to the whole system. The LFC system was already implemented in JaCaMo, which has been used by several winning teams in past editions of the contest. Our goal in this work was to verify a specific part of the system; the map merge protocol. Both of these things contributed to our exploration of using a \gls{csp} specification of the protocol. However, this does not preclude its integration with other types of formal methods applied to the LFC system, which can provide greater confidence in the correctness of the system (as well as guiding the development of new functions)~\cite{FarrellLF:IFM018}.

\gls{csp} has been used in other approaches for multi-agent systems. Examples include an approach that combines a \gls{csp} encoding of agent communications with a first-order logic framework~\cite{izumi1990csp}; a \gls{csp} framework for a Java-based ``cognitive agent architecture'' called Cougar~\cite{gracanin2005csp}, where the model is used to verify properties about the code generated from the Cougar system; and a timed \gls{csp} model of a multi-agent manufacturing system~\cite{yeung2011behavioral}. However, each of these approaches is (like ours) specific to its example application. 

Another approach~\cite{kacem_formal_2007} for multi-agent systems that involves \gls{csp} presents a translation from CSP-Z (a combination of \gls{csp} and Z~\cite{spivey_z_1992}) to Promela, the input language of the SPIN~\cite{holzmann_model_1997} model checker. This translation appears to be needed to side-step some inadequacy with a previous version of \gls{fdr}. They demonstrate their approach using a CSP-Z specification of an air traffic control system. Our work makes use of `pure' \gls{csp}, and doesn't require the specification to be translated into a different language for model checking, so we can be more confident of our results. However, updating this approach for the current versions of \gls{fdr} and SPIN could be useful if the protocol had specification temporal properties that needed checking.

Finally, there is work on the Agent Communication Programming Language (ACPL)~\cite{vanEijk2003}, which is a process algebra that takes some inspiration from \gls{csp}'s approach to concurrency to model the basics of agent communication. ACPL was also used as the basis for a formal compositional verification framework for agent communication~\cite{van_eijk_verification_2003}. While the map merge protocol tackled in our work does use agent communication, we are verifying the protocol not the communication itself.

Other process algebras have been used to specify and verify multi-agent systems~\cite{luckcuck_formal_2019}. For example, in~\cite{Akhtar2014} the process algebra \gls{fsp} and \gls{piadl} are used to specify safety and liveness properties for a multi-agent system, The multi-agent program is checked for satisfaction of these properties, as is the agent architecture (which is written in \gls{piadl}). 
Looking further afield, process algebras have been applied to similarly distributed, cooperative systems. For example, the Bio-PEPA process algebra has been used to model robot swarms~\cite{massink2013use}, specifying behaviour that enables the swarm to perform a foraging task. They found that their approach enabled a wider range of analysis methods, when compared to other modelling approaches.

\section{Conclusion}
\label{sec:conc}

This paper describes the application of formal specification and verification techniques (in \gls{csp}) to an existing communication protocol used to merge maps in a multi-agent system. The protocol was used by the LFC team in the MAPC 2019. This work provides extra confidence that the LFC team's map merging protocol works correctly, which was difficult to check using testing alone. The work also explores the utility of \gls{csp} for modelling multi-agent systems.

The merge protocol is critical to the performance of the multi-agent system, all of the information needed for the agents to participate effectively in the competition is stored in the agent's maps. Without a coherent map, the agents would not have been able to cooperate to achieve their mission. 

Using the model checker \gls{fdr}, our \gls{csp} specification of the protocol was validated (through checking that it could perform traces representing scenarios drawn from the MAPC 2019) and verified to be free of divergences, and non-determinism, and that it could eventually merge all the maps without deadlocking.  We conclude that \gls{csp}'s focus on concurrent communicating systems makes it well suited to specifying this kind of communications protocol. 

Although the merge protocol is the most complex communication protocol used in the LFC system, other behaviours also require some form of validation and verification. The identification process (mentioned in Sect.~\ref{sec:mergeProtocol}) could be specified in \gls{csp} either as an addition to the specification presented in this paper, or separately. \gls{csp} is useful for modelling concurrent communication, but there may be other formal method techniques that are more appropriate for the remaining behaviours. As indicated by the results in~\cite{Cardoso20a}, different parts of the system may require distinct verification techniques. The use of \gls{csp} for modelling agent interaction protocols that make use of the interaction dimension in JaCaMo also requires further investigation. These are left for future work.

\bibliographystyle{splncs04}
\bibliography{emas}
%


\end{document}